\documentclass{aa}  

\usepackage{graphicx}
\usepackage{txfonts}
\usepackage{lipsum}
\usepackage{subcaption}         
\usepackage{lscape}             
\usepackage{placeins}          

\usepackage[dvipsnames]{xcolor}
\usepackage[normalem]{ulem}                

\def\Ne{{N_{e}}}

\begin{document}

   \title{Solar models using OPLIB monochromatic opacities}



   \author{M. Vidal \inst{1}
        \and M. Deal\inst{1}
        \and O. Richard\inst{1}
        \and Y. Lebreton\inst{2,3}
        \and J. Wang\inst{1}
        \and J. Colgan\inst{4}
        \and C.J. Fontes\inst{4}}

   \institute{LUPM, Universit\'e de Montpellier, CNRS, Place Eug\`ene Bataillon, 34095 Montpellier, France\\
             \email{marin.vidal@umontpellier.fr}
             \and LIRA, Observatoire de Paris, Université PSL, Sorbonne Université, Université Paris Cité, CY Cergy Paris Université, CNRS, 92190, Meudon, France
             \and Univ Rennes, CNRS, IPR (Institut de Physique de Rennes) - UMR 6251, F-35000 Rennes, France
             \and Los Alamos National Laboratory, Los Alamos, NM 87545, USA}

   \date{Received October 19, 2026}

 
  \abstract
   {Opacity is a key ingredient in describing the transport of energy in stars. There are several methods for calculating stellar opacities, resulting in different tables (OPAL, OP, OPAS, and OPLIB). These opacities are available in two distinct forms: tables that regroup the Rosseland mean opacity pre-calculated at predefined chemical mixtures, and monochromatic opacity spectra of individual chemical elements, which allow for the computation of Rosseland mean opacity for a precise mixture.}
   {This study presents our implementation of the OPLIB monochromatic opacity data in Cesam2k20 and compares the solar-calibrated model obtained with these opacities to helioseismic data in order to evaluate the improvements made by using monochromatic opacities.} 
   {We implemented the OPLIB monochromatic opacity data in the Cesam2k20 stellar evolution code following the method already used for the OP monochromatic opacity data.}
   {We found that calibrated solar models produced with monochromatic opacity data (OP and OPLIB) agree better overall with helioseismic constraints than those produced with pre-calculated opacity tables.}
   {}


   \maketitle
   \nolinenumbers

\section{Introduction}
    The Rosseland mean opacity (RMO) is a key ingredient in describing the internal structure of stars because it is involved in the energy transport equation and in many other areas. It is also one of the most difficult components to determine accurately because it depends on several stellar parameters, in particular, on the internal distribution of chemical elements. However, there are several approaches to compute opacities, and for many years, intense efforts have been made to produce complete and accurate opacity tables that can be used in stellar evolution codes: OPAL \citep{iglesias1996updated}, OP \citep{seaton2005opacity,badnell2005updated}, OPAS \citep{blancard2012solar}, OPLIB \citep{colgan2016new}, and SCO-RCG \citep{pain2015accounting}. They are classed into two categories: the first category concerns tables containing Rosseland mean opacities calculated for predefined chemical compositions (type 1, varying the hydrogen, helium, and metal mass fractions; and type 2, also varying carbon and oxygen), while the second category concerns monochromatic opacities. The Rosseland mean opacity tables are adapted when the mixture of metals does not vary with the evolution (except for C and O for type 2 opacity tables during the helium-burning phase) \cite[see e.g.][for stars with masses larger than 2~M$_\odot$]{huibonhoa21}. However, the metal mixture can be strongly affected, for example when the effect of atomic diffusion is strong due to radiative accelerations \citep{richer2000,richard01,deal16,huibonhoa2018} or in the presence of accretion of heavy elements \citep[][and references therein]{deal15}. In this case, the monochromatic opacity data are needed to compute self-consistent Rosseland mean opacities in any inner shell at any evolutionary point. While OPAL monochromatic opacity data are not public and only available within the Montpellier-Montréal stellar evolution code \citep{turcotte98}, OP monochromatic opacity data are public and implemented in several stellar evolution codes (e.g. TGEC \citealt{huibonhoa21}, MESA \citealt{paxton2015}, Cesam2k20 \citealt{deal18}, and YREC \citealt{pinsonneault2026YREC}). The OPLIB monochromatic opacity data \citep{colgan2016new}, computed with the ATOMIC code, have been made available recently. These OPLIB monochromatic opacity data have never been used in stellar evolution codes. We present the implementation of the OPLIB monochromatic opacity data in Cesam2k20 and their impacts on the solar models. Section~\ref{sec:stellar_opacity} provides an overview of the original structure of OPLIB monochromatic opacity data \citep{colgan2016new} and explains the method we used to construct new opacity  tables for integration into stellar evolution codes. Section~\ref{sec:implementation} shows the implementation of these adapted data in the Cesam2k20 code \citep{morel1997,morel2008,marques2013,manchon2026} and compares them with previous theoretical work and available measurements. Finally, Sect.~\ref{sec:solar_model} presents the solar model calculated using these opacities and analyses its compatibility with helioseismic constraints. A conclusion is provided in Sect.~\ref{sec:conclusion}.
\section{Stellar opacity and OPLIB monochromatic data}
    \label{sec:stellar_opacity}
    \subsection{Stellar opacity}
        In stellar evolution codes, the Rosseland mean opacity is the most appropriate mean opacity for the deep interior, in contrast to the Planck mean opacity, which is suitable for an optically thin medium \citep{huebner2014opacity}. By definition, the RMO $\kappa_{R}$ is a harmonic mean of the total monochromatic opacities $\kappa_{\nu}^{tot}$ weighted by the temperature derivative of the Planck function $B_{\nu}$,
        \begin{equation}
            \frac{1}{\kappa_{R}}=\frac{\int_{0}^{\infty}\frac{1}{\kappa_{\nu}^{tot}}\frac{\partial B_{\nu}}{\partial T}d\nu}{\int_{0}^{\infty}\frac{\partial B_{\nu}}{\partial T}d\nu}
            \label{eq:opa_ross_1}.
        \end{equation}
        Equation~\ref{eq:opa_ross_1} can be simply rewritten as \citep{badnell2005updated} 
        \begin{equation}
            \frac{1}{\kappa_{R}}= \int_{0}^{\infty}\frac{F(u)}{\kappa_{\nu}^{tot}(u)}du
            \label{eq:opa_ross_2},
        \end{equation}
        where $u$ is the dimensionless photon energy, and $F(u)$ is the normalised temperature derivative of the Planck function,
        \begin{equation}
            u = \frac{h\nu}{k_BT} \quad \text{;} \quad F(u)= \frac{15}{4\pi^4}\frac{u^4 e^{-u}}{(1-e^{-u})^2}
            \label{eq:F_u},
        \end{equation}
        where $h$ is Planck's constant, and $k_{B}$ is Boltzmann's constant. 
        \paragraph{}
        The total monochromatic opacity $\kappa_{\nu}^{tot}$ of a mixture at the frequency $\nu$ in Eq.\ref{eq:opa_ross_1} is the combination of the total scattering opacity $\kappa_{\nu,tot}^{scatt}$ and the total  absorption opacity $\kappa_{\nu,tot}^{abs}$ ,
        \begin{equation}
            \label{eq:kappa_tot1}
            \kappa_{\nu}^{tot} = \kappa_{\nu,tot}^{abs} + \kappa_{\nu,tot}^{scatt}.
        \end{equation}
        The absorption opacity of the element $k$ is the sum of the inverse Bremsstrahlung $\kappa_{\nu,k}^{ff}$ (free-free), photo-ionisation $\kappa_{\nu,k}^{bf}$ (bound–free), and photo-excitation $\kappa_{\nu,k}^{bb}$ (bound–bound) contributions,  
        \begin{equation}
            \kappa_{\nu,k}^{abs}= \left ( 1-e^{-u} \right ) \left( \kappa_{\nu,k}^{ff} + \kappa_{\nu,k}^{bf} + \kappa_{\nu,k}^{bb} \right )
            \label{eq:opa_abs}.
        \end{equation}
        The factor in front of the absorption contribution $(1-e^{-u})$ is a correction factor for stimulated emission. The scattering opacity is the sum of the Compton- and Rayleigh-scattering contributions.

    \subsection{OPLIB database}
        \label{sec:oplib_data}
        \begin{figure}
           \centering
           \includegraphics[width=\hsize]{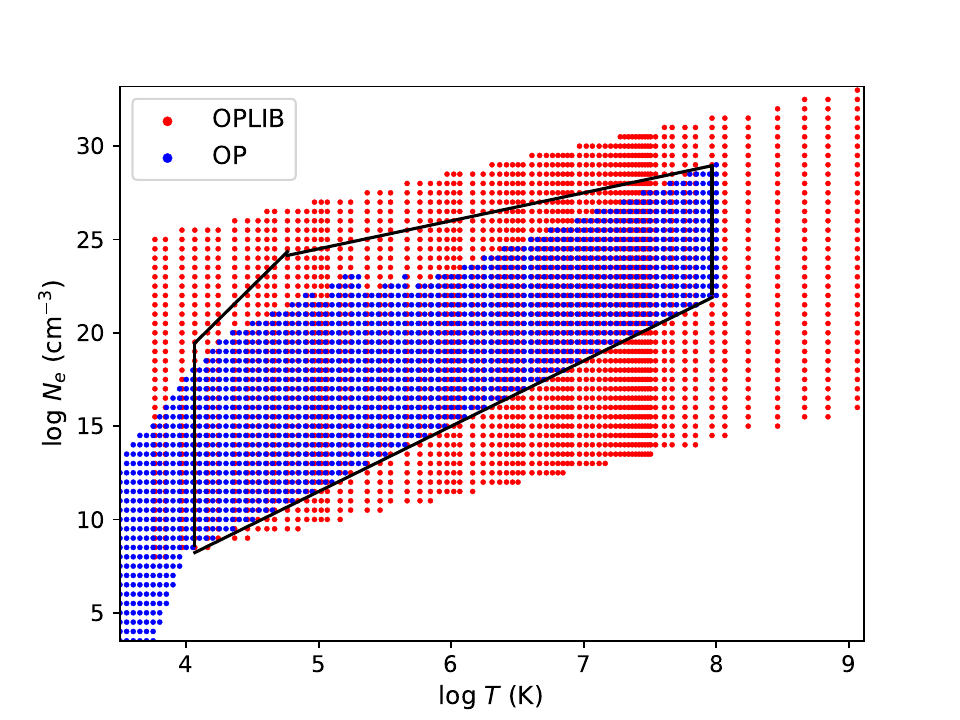}
           \caption{Grid point comparison between OP data \citep{seaton2005opacity} in blue and the OPLIB data downloaded in this work in red. The area within the black lines represents our working grid.}
           \label{fig:data_grid_mono}
        \end{figure}
        The OPLIB\footnote{Opacities available at http://aphysics2.lanl.gov/opacity/lanl/} opacities are computed using the ATOMIC code \citep{magee200414th,hakel2006new,fontes2015alamos}, which can be run in LTE or non-LTE mode. The equation-of-state (EOS) model used in ATOMIC, known as ChemEOS, is based on the minimisation of the free energy in the chemical picture and appears to be a reasonable and robust approach to determining atomic state populations over a wide range of temperatures and densities. The OPLIB database can produce monochromatic opacity spectra (total, absorption, and scattering) on a large photon energy grid $u$ (from $10^{-4}$ to $30 000$) for elements with atomic numbers $Z=1-30$ and for a wide range of temperatures (from $0.5$ eV to $100$ keV)\footnote{1 keV corresponds to $T=1.1604\times10^{7}$ K} and mass densities (for more details about how OPLIB opacities are computed, see \cite{colgan2016new}).
        \paragraph{}
        By using a modified version of the pyTOPScrape\footnote{https://github.com/tboudreaux/pytopsscrape} script \cite[first introduced in][]{boudreaux2023updated}, we queried the database and retrieved the OPLIB monochromatic opacity data for all elements (from hydrogen to zinc) and available temperatures and for all available electron densities with the same grid stepping as the OP tables ($\Delta\log\Ne=0.5$, see \citealt{seaton2005opacity}). It is a better choice to classify the monochromatic opacity data using a $T-\Ne$ grid instead of a $T-\rho$ grid because it is equivalent to using a $T-\eta$ grid, where $\eta$ is the electron degeneracy parameter (see Eq.~\ref{eq:eta}), which is more suitable for creating an atomic opacity library \cite[see Chap.10 of][]{huebner2014opacity}. Equation~\ref{eq:kappa_tot1} then becomes
        \begin{equation}
            \label{eq:kappa_tot}
            \begin{split}
                \kappa_{\nu}^{tot}(T,\eta)&=\kappa_{\nu,tot}^{abs}(T,\eta) + \kappa_{\nu,tot}^{scatt}(T,\eta) \\
                &=\sum_k X_k \kappa_{\nu,k}^{abs}(T,\eta) + \kappa_{\nu,tot}^{scatt}(T,\eta)
            \end{split},
        \end{equation}
        where $X_{k}$ is the mass fraction of element $k$. This equation is valid under the assumption of equilibrium between the electron temperature $T$ and electron degeneracy $\eta$.
        \paragraph{}
        One advantage of working in the $T-\Ne$ plane is that it is common to all elements, unlike the $T-\rho$ plane \citep{colgan2016new}. It is also the way in which OP tables are provided, which facilitates the comparison between the two datasets. Figure~\ref{fig:data_grid_mono} shows the downloaded OPLIB data grid, which covers a large part of the $T-\Ne$ plane, along with comparison points from the OP grid. The range of electron densities in the OPLIB grid increases proportionally as the temperature increases because the OPLIB opacities are computed on a grid of constant electron degeneracies ($\eta$). We also remark compared to the OP grid that the density of points in the OPLIB grid is lower at low temperatures. The temperature grid of OPLIB varies, and the OP grid covers a larger area at low temperature. The choice of electron temperatures in the OPLIB tables was made to maximise the coverage of explicitly computed points in key regions of the opacity tables in an effort to minimise interpolation errors while also providing a comprehensive set of opacities across a very wide range of temperatures.
        
\begin{figure}
           \centering
           \includegraphics[width=\hsize]{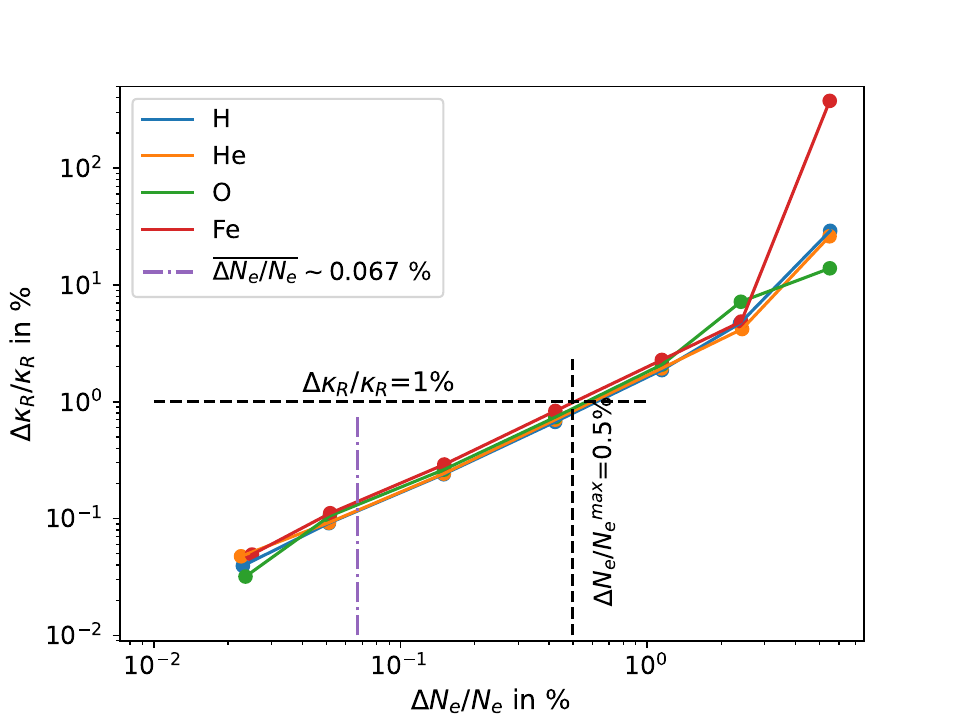}
           \caption{Average relative difference of RMO (in $\%$) as a function of the precision on the electron density $\Delta{\Ne/\Ne}$ (in $\%$) for hydrogen, helium, oxygen, and iron.}
           \label{fig:delta_Ne_preci}
        \end{figure}

        \paragraph{}
        On the OPLIB website, the user can query the database by temperature and mass density instead of temperature and electron density. We inferred the mass density, corresponding to the chosen electron density, for element $k$ by iterating over the opacities from the OPLIB website and using the following relation:
        \begin{equation}
            \Ne(\rho) = \frac{\overline{N}_{f,k}\times \rho}{\mu_{k}}
            \label{eq:Ne_rho},
        \end{equation}
        where $\mu_{k}$ is the atomic mass, and $\overline{N}_{f,k}$ is the average number of free electrons per ion for element $k$. We validated the mass density, $\rho$, when we retrieved the chosen electron density with a relative deviation $\Delta{\Ne/\Ne}\leq0.5\%$. This criterion depends very strongly on the number of significant digits of $\overline{N}_{f,k}$ given by the database. Fig.~\ref{fig:delta_Ne_preci} shows the maximum relative difference of the RMO (in $\%$) as a function of the relative deviation $\Delta{\Ne/\Ne}$ for three single elements. We decided to impose $\Delta{\Ne/\Ne}\leq0.5\%$ to optimise the computation time and to retain adequate precision on the RMO, which then remains below 1\%. The mean value of $\Delta{\Ne/\Ne}$ is about $0.067\%$ for all the points we considered for the comparison with OP data, leading to a precision on the opacity of about $0.3\%$ on average. 
        
        \paragraph{}
        We only retrieved the monochromatic absorption spectra from the website since the scattering spectra of pure elements cannot be added together because of collective effects \citep{boercker1987collective,kilcrease2001plasma}. Each spectrum was retrieved for values of dimensionless photon energies $u$ between $10^{-4}$ and 30. This choice was made since the relative difference between the integration on all photon energies ($u \in [10^{-4}, 30 000]$) and on the chosen values ($u \in [10^{-4}, 30]$) is lower than 0.015$\%$. Furthermore, integrating over only a portion of the frequencies optimises computation time since the integration must be made at each time step and for each mesh point of the model. Finally,  when we integrate total monochromatic opacities in the following sections, we use Eq.\ref{eq:opa_ross_2} which is equivalent to fixing the refractive index $n_{\nu}$ to 1 in the equation presented in \cite{colgan2016new}.
    \subsection{Low-temperature opacities}
        \label{section:low_temp}
        The OP and OPLIB monochromatic opacity data can be used to compute high-temperature RMO, but they need to be complemented at low temperatures by specific tables. When the users decide to compute a model with monochromatic opacity data (OP and OPLIB) in Cesam2k20, the following method is used:
        \begin{itemize}
            \item For $T>10^4$~K: RMOs are computed from monochromatic data.
            \item For $T\leq10^4$~K: RMOs are computed using pre-calculated type 1 opacity tables from the same source as the monochromatic data (OP and OPLIB) complemented with the Wichita tables provided for the same solar mixture (see Sect.\ref{section:solar_model}; \citealt{ferguson2005low}). High- and low-temperature tables are merged with a linear transition around $T=10^{4}$K in the range $\log T\in [3.95, 4.05]$.
        \end{itemize}
        For models using only pre-calculated type 1 RMO tables, the treatment at low temperature is the same as in the procedure described above.
        Negligible differences were found in the Hertzsprung-Russell diagram when comparing models using \cite{ferguson2005low} and more recent updated low-temperature tables such as \cite{marigo2022}, as shown in \cite{reeve2023}.
        For solar models, the small differences are compensated for by a variation in the mixing length parameter $\alpha_{\text{MLT}}$ in the calibration process. In the following, we restrain the analysis and discussions to the area outlined in black in Fig.~\ref{fig:data_grid_mono}, that is, the area related to pure OPLIB tables.
        
\section{Implementation in a stellar evolution code}
    \label{sec:implementation}
    To implement and use OPLIB monochromatic opacity data in Cesam2k20, we followed the processing method already available for OP monochromatic opacity data \citep{deal18} including the optimisation proposed by \cite{huibonhoa21}. However, some modifications were made, particularly to compute scattering and interpolation.
    \subsection{Scattering contribution}
        \label{sec:scatt}
        To compute the scattering opacity, we followed the method presented in \cite{colgan2016new}. Photon scattering is the combination of Rayleigh scattering and Compton scattering. For Rayleigh scattering, we only calculated the contribution of hydrogen and helium because the database does not provide the ion fractions at each point of the grid for other elements. However, the Rayleigh-scattering contribution dominates at low temperatures \citep{colgan2016new}, close to the temperature domain where we use pre-computed tables (see Sect. \ref{section:low_temp}), which means that this contribution does not affect the computation of the total scattering opacity. The Rayleigh scattering is also strong at low photon energies, where the free-free or bound-bound contributions are much more prominent.
        \paragraph{}
        To compute the Compton scattering, we followed the method adopted for OPLIB data  \citep{kilcrease2001plasma} using some routines based on \citealt{boercker1987collective} for OP monochromatic data, adapted to Cesam2k20 by \cite{deal18}. To be more specific, the Compton-scattering values come from Sampson tables \citep{sampson1959opacity}, but were modified to include collective and Pauli blocking effects. An important point to emphasise is that we have not yet included the relativistic correction for these effects. More details are given in Appendix \ref{appendix:scatt}.
        \paragraph{}
        \begin{figure}
           \centering
           \includegraphics[width=0.9\hsize]{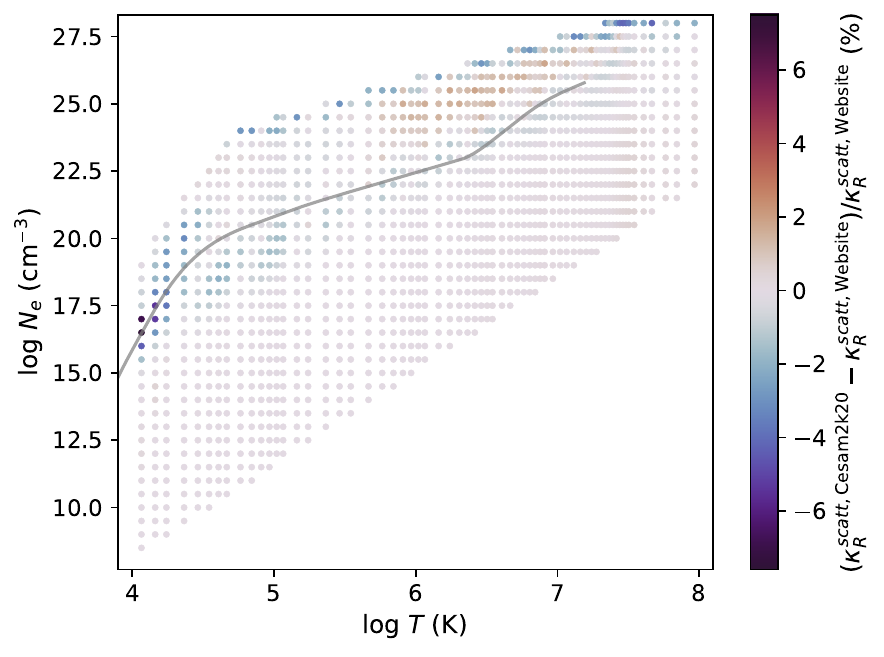}
           \caption{Relative difference of the RMO computed only with the scattering contribution. The grey line corresponds to a solar model computed with AAG21 mixture.}
           \label{fig:diff_scatt}
        \end{figure}
        Figure~\ref{fig:diff_scatt} shows the relative differences (in percent) between the RMO computed using only the scattering contribution from our computation ($\kappa_{R}^{scatt,\rm{Cesam2k20}}$) and that obtained from the OPLIB website ($\kappa_{R}^{scatt,\rm{Website}}$). The chosen mixture is that of the base of the convective zone of a solar calibrated model (model OPLIB$_{T}^{\rm{AAG21}}$ of Table~\ref{tab:list models}). The differences are between $-7.5\%$ and $+7.5\%$. We can distinguish three regions in the $T-\Ne$ plane. The differences at low temperature ($\log T \sim4.25-4.75$) and at density $\log \Ne \sim16-18$ are related to our computation method of the collective effect factor $f(\eta,\delta)$ (see Appendix~\ref{appendix:scatt}). Following \cite{boercker1987collective}, our collective effect factors were computed in the non-relativistic case, unlike those from OPLIB data \citep{kilcrease2001plasma}, and this matters in this region. There are also differences due to electronic density values: as already mentioned, we searched for a mass density corresponding to a specific electron density with a small relative deviation. In this specific region, this $\rho-\Ne$ correspondence is different between that of Cesam2k20 and that of the website.
        \paragraph{}
        In the second region at $\log T\sim6.5$ and $\log \Ne \sim24$, the difference of $\sim2\%$ results from our computation method of the scattering opacities for metals. We did not include the Rayleigh scattering for these elements, and we approximate the number density of the free electron $n_{e}$ in Eq.~\ref{eq:compton_scatt} to the total electron number density. Without the information of the population of the ground and ionised states of these elements for each point of the grid, we cannot do otherwise.
        Finally, the $4\%$ differences at high density aris because we neglected the relativistic Pauli blocking factor (see \citealp{kilcrease2001plasma}). However, these differences are negligible for solar models (grey line in Fig.~\ref{fig:diff_scatt}).
    \subsection{$\kappa_{R}$ calculation and interpolation method}
        \label{section:interpolation}
        During the computation of stellar structure models, the RMO $\kappa_{R}$ of each layer at position $r$ in the star depends on the local temperature $T$, electron density $\Ne$, and chemical mixture. To obtain it, we used the computation method for the OP monochromatic data that is already implemented in Cesam2k20 \citep{deal18}. This method consists of computing the 16 surrounding RMO values in the temperature-electron density plane $\log T -\log \Ne$ (four in each direction) and interpolating among these values. We computed the RMO value at each of these 16 points as follows :
        \begin{itemize}
            \item firstly, we computed the total absorption opacity $\kappa_{\nu,tot}^{abs}$ of the mixture with all absorption opacity data of each element $k$ weighted by their mass fraction $X_{k}$;
            \item secondly, we computed the total opacity $\kappa_{\nu}^{tot}$ using $\kappa_{\nu,tot}^{abs}$ and $\kappa_{\nu,tot}^{scatt}$ by using Eq. \ref{eq:kappa_tot} and the scattering contribution presented in Sect.~\ref{sec:scatt};
            \item finally, we  obtained $\kappa_{R}(T_{e},\Ne,X_{k})$ by integrating $\kappa_{\nu}^{tot}$ on the frequencies with Eq.~\ref{eq:opa_ross_2} for the first 12200 points ($10^{-4}\leq u \leq 30$).
        \end{itemize}
        
        These computation steps are necessary, even though they are expensive in resource and computation time, in particular, for the calculation of the integral over frequencies \citep{huibonhoa21}. The computation time required to obtain a model using OPLIB monochromatic data is 30 times greater than that for a model using the OPLIB pre-calculated table, and the requirements in memory is increased by at least the size of the data ($\sim 4$ GB).
        \paragraph{}
        We modified the OP interpolation function to better adapt it to the OPLIB data: the OPLIB temperature-grid\footnote{see https://aphysics2.lanl.gov/static/opacdocs/opactemps.html} is irregularly spaced in temperature, unlike the OP grid. The interpolation function was changed from bi-cubic no-spline to bi-cubic spline functions. We also added an optional parameter that determines how $\Ne$ is obtained: either from the average number of free electrons derived from the OPLIB database or from ion fractions using the Saha equation. We chose the Saha equation to calculate $\Ne$ by default to be consistent with the rest of the Cesam2k20 code. In the remainder of this paper, the models that include OPLIB monochromatic opacity data are computed with this option (except for the model OPLIB$_{M}^{\rm{AAG21,}\rho}$ presented in Fig.~\ref{fig:relat_diff_opal_oplib}).
    \subsection{Validation of the $\kappa_{R}$ computation method}
        \label{section:verif}
\begin{figure}
           \centering
           \includegraphics[width=0.9\hsize]{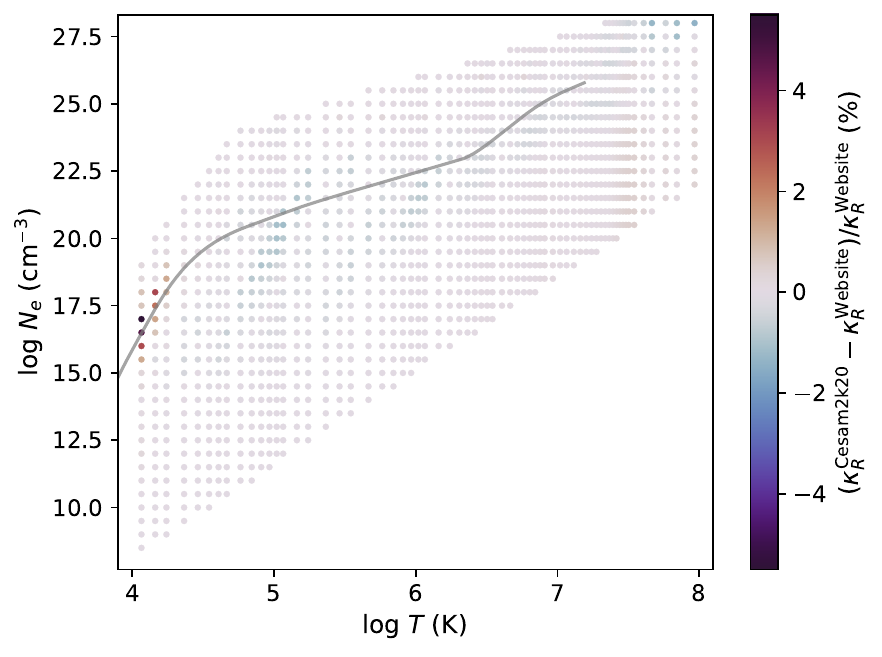}
           \caption{Relative difference between the values of $\kappa_{R}^{\rm{Website}}$ from the OPLIB website and $\kappa_{R}^{\rm{Cesam2k20}}$ computed in Cesam2k20 with the mixture of the OPLIB$_{T}^{\rm{AAG21}}$ at the convective zone base. The difference values are included between $-5.5\%$ and $+5.5\%$.}
           \label{fig:diff_tot_abs_compton}
        \end{figure}

        Figure~\ref{fig:diff_tot_abs_compton} shows the relative difference between the RMO $\kappa_{R}^{\rm{Website}}$ for all $T-\Ne$ points of our working grid retrieved from OPLIB website and the RMO $\kappa_{R}^{\rm{Cesam2k20}}$ computed in Cesam2k20 with the method presented in the previous section. The metal mixture is that at the convective zone base of the model OPLIB$_{T}^{\rm{AAG21}}$ presented in Table~\ref{tab:list models}.
        \paragraph{}
        The maximum differences are seen at low temperatures ($\log T <4.25$) and at density $\log \Ne <16$. Unlike in Fig.~\ref{fig:diff_scatt}, these differences are linked to the computation of the total absorption opacities $\kappa_{\nu,tot}^{abs}$, which contribute most to the RMO for these temperature and density values. As we already mentioned (see Sect.~\ref{section:interpolation}), the total absorption opacity of a mixture $\kappa_{\nu,tot}^{abs}$ is the sum of all pure absorption opacities $\kappa_{\nu,k}^{abs}$ of each element $k$ in the mixture at the same temperature $T$ and electron density values. Here, these pure absorption opacities are sometimes combined at different $\Ne$, due to their precision (see Sect.~\ref{sec:oplib_data}) and because the RMO value, computed by only absorption opacities, varies greatly with only a small variation in $\Ne$ in this region, this causes the differences shown in Fig.~\ref{fig:diff_tot_abs_compton}. However, only differences between high-temperature opacity sources above $\log T > 4.3$ can affect the model evolution \citep[see][]{boudreaux2023updated}.
        \paragraph{}
        The other differences in Fig.~\ref{fig:diff_tot_abs_compton} are the same as in Fig.~\ref{fig:diff_scatt} and are due to our scattering computation method, that is, the computation for non-relativistic cases and approximation for metals. We note that the differences are smaller than before since the absorption contribution attenuates it, especially for the scattering for the metals, where in this region absorption opacities contribute $99\%$ to the RMO.
       
\section{Solar models}
    \label{sec:solar_model}
    \subsection{Standard solar model calibration}
        \label{section:solar_model}
        To go even further in validating our implementation, we calculated calibrated solar models using OPLIB monochromatic data and compared our results with helioseismic inference of the sound speed and density profiles from BiSON-13 oscillation frequencies by \cite{basu2009}. We used Cesam2k20 to compute solar-calibrated models (SCM) by adjusting the initial helium mass fraction $Y_{0}$, the initial metallicity $Z_{0}$, and the mixing length parameter $\alpha_{\rm{MLT}}$ to obtain the solar luminosity $L_{\odot}$, the solar radius $R_{\odot}$, and the photospheric abundance $(Z/X)_{\rm{surf}}$  at the current solar age. The set of observational constraints is listed in Table \ref{tab:constraints}.
        \begin{table}
        \caption{Values of observational constraints used to calibrate each solar-calibrated models.}
        \label{tab:constraints}
        \centering
        \renewcommand{\arraystretch}{1.35}
        \begin{tabular}{c|cc}
         \hline
         Parameter & Notation & Value\\
         \hline \hline
         Age & Age & 4570 [Myrs] \\
         Luminosity & $L_{\odot}$ & $3.846\times10^{33}$ [erg.s$^{-1}$] \tablefootmark{a}\\
         Radius & $R_{\odot}$ & $6.9599\times10^{10}$ [cm] \tablefootmark{a}\\
         Photospheric & $(Z/X)_{\rm{surf}}$ & 0.0187 (AAG21)\\
         abundance& & 0.02250 (MB22)\\
        \hline
        \end{tabular}
        \tablefoot{\\ \tablefoottext{a}{Constants are taken from IAU 2018 \citep{mamajek2015}}}
        \end{table}
        
        Except for the choice of solar mixture and opacity source, all SCMs were computed with the same input physics. We used the EOS OPAL 2005 \citep{rogers2002}\footnote{The EOS used to compute OP and OPLIB opacities are not available, which prevents a fully consistent computation of stellar models for these aspects.}. The temperature gradient in the convective zones was computed with the mixing length theory (MLT). The mixing length $l$ is defined as $l = \alpha_{\rm{MLT}}H_{p}$ where $\alpha_{\rm{MLT}}$ is the convection parameter, and $H_p$ is the pressure scale height \citep{bohm1958}. Along the evolution, we took a constant $\alpha_{\rm{MLT}}$. The thermonuclear reactions we used are the combination of PP chain and CNO cycle for 25 elements. The rates of these reactions were taken from NACRE \citep{aikawa2006} with the rate from LUNA for $\mathrm{^{14}N (p, \gamma)^{15}O}$ \citep{broggini2018}. The atmosphere was retrieved from the Hopf fully radiative $T(\tau)$ law \cite[more details are given in][]{manchon2025}. We applied microscopic diffusion of chemical elements using the \cite{michaud93} formalism. There was no additional turbulent diffusion coefficient, rotation, or mass loss. The models were computed starting at the pre-main-sequence.

        \begin{table}
        \caption{List of the solar-calibrated models.}
        \label{tab:list models}
        \centering
        \renewcommand{\arraystretch}{1.6}
        \resizebox{0.5\textwidth}{!}{
        \begin{tabular}{c|cc}
         \hline
         Opacity Source & \citealt{asplund2021chemical} & \citealt{magg2022} \\
         \hline \hline
         \citealt{iglesias1996updated}& OPAL$_{T}^{\rm{AAG21}}$& OPAL$_{T}^{\rm{MB22}}$ \\
         \citealt{seaton2005opacity}& OP$_{T}^{\rm{AAG21}}$& OP$_{T}^{\rm{MB22}}$ \\
         \citealt{seaton2005opacity}& OP$_{M}^{\rm{AAG21}}$& OP$_{M}^{\rm{MB22}}$ \\
         \citealt{farag2024expanded}& OPLIB$_{T}^{\rm{AAG21}}$& OPLIB$_{T}^{\rm{MB22}}$ \\
         \citealt{colgan2016new} & OPLIB$_{M}^{\rm{AAG21}}$& OPLIB$_{M}^{\rm{MB22}}$ \\
        \hline
        \end{tabular}
        }
        \tablefoot{\\ The subscript in the name of the models represents the choice of using pre-calculated opacity tables ($_{T}$) or monochromatic opacity data ($_{M}$). The superscript in the name of the models indicates the choice of the mixture (AAG21 or MB22).}
        \end{table}
        We performed the calibrations for two photospheric solar mixtures, AAG21 \citep{asplund2021chemical} and MB22 \citep{magg2022}, and for five different opacity sources. The names of all SCMs we computed are listed in Table~\ref{tab:list models}.
        \paragraph{}
        \begin{table}
            \caption{Optimal parameters found for solar-calibrated models.}
            \label{tab:optimal}
            \centering
            \renewcommand{\arraystretch}{1.35}
            \begin{tabular}{l|cccc}
                 \hline
                 Models & $Y_{0}$ & $Z_{0}$ & ($Z/X$)$_{0}$ & $\alpha_{\rm{MLT}}$\\
                 \hline \hline 
                 OPAL$_{T}^{\rm{AAG21}}$&0.2630 &0.0154 &0.0214&1.77705 \\
                 OP$_{T}^{\rm{AAG21}}$&0.2616&0.0154 &0.0213 &1.79214\\
                 OP$_{M}^{\rm{AAG21}}$&0.2613 &0.0154 &0.0213 &1.79411\\
                 OPLIB$_{T}^{\rm{AAG21}}$&0.2528&0.0155 &0.0213 &1.82244 \\
                 OPLIB$_{M}^{\rm{AAG21}}$&0.2556&0.0156 &0.0214 &1.82541 \\
                 \hline
                 OPAL$_{T}^{\rm{MB22}}$&0.2725 &0.0182 &0.0256 &1.80454 \\
                 OP$_{T}^{\rm{MB22}}$&0.2718&0.0182 &0.0256 &1.81407\\
                 OP$_{M}^{\rm{MB22}}$&0.2697 &0.0181 &0.0255 &1.82791\\
                 OPLIB$_{T}^{\rm{MB22}}$&0.2673&0.0182 &0.0255 &1.84124 \\
                 OPLIB$_{M}^{\rm{MB22}}$&0.2653&0.0184 &0.0256 &1.85388 \\
                 \hline
            \end{tabular}
        \end{table}
        
        The calibration parameters of these SCMs are presented in Table~\ref{tab:optimal}. The OP$_{T}$ and OPLIB$_{T}$ models have an initial mass fraction $Y_{0}$ and $Z_{0}$ for both mixtures, which are consistent with previous studies \cite[e.g.][and references therein]{farag2024expanded}. For each mixture, the OP$_{M}$ models have similar initial parameters with OP$_{T}$. The same results are found between the OPLIB$_{M}$ and OPLIB$_{T}$ models with a difference of $1.1\%$ for the Y$_{0}$ for AAG21, however. This difference affects the final helium mass fraction at the base of the convective zone ($Y_{\rm{bcz}}$). Overall, the OPLIB models have a lower initial helium mass fraction than OPAL and OP models.
    
    \subsection{Comparison between models using pre-calculated opacity tables and monochromatic data}
        \label{sec:comparison_opacity}
\begin{figure}
            \includegraphics[width=0.9\hsize]{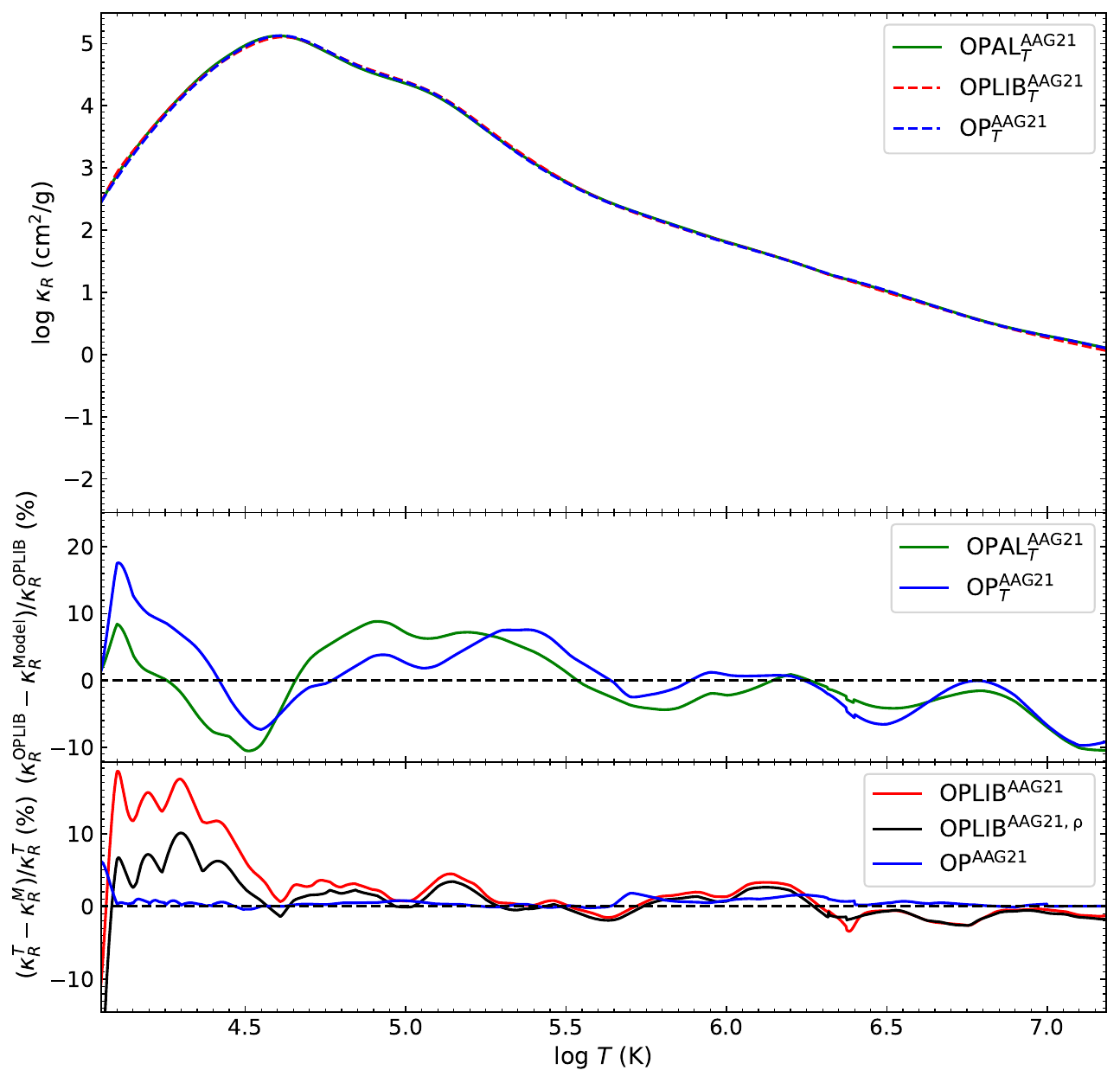}
            \caption{Rosseland mean opacity profile in OPAL$^{\rm{AAG21}}_{T}$, OP$^{\rm{AAG21}}_{T}$, and OPLIB$^{\rm{AAG21}}_{T}$ models (top). We show the relative difference of RMO between OPLIB$_{T}^{\rm{AAG21}}$ model and both OP$_{T}^{\rm{AAG21}}$ and OPAL$_{T}^{\rm{AAG21}}$ models (middle) and the relative differences between pre-calculated tables ($\kappa_{R}^{T}$) and monochromatic opacity data ($\kappa_{R}^{M}$) for OP (blue curve) and OPLIB (red and black curves) opacities (bottom). The black line represents the difference between OPLIB$_{T}^{\rm{AAG21}}$ and OPLIB$_{M}^{\rm{AAG21,}\rho}$, a model with the same internal structure as OPLIB$_{M}^{\rm{AAG21}}$, but using the ion fractions derived from the OPLIB database to compute $\Ne$.}
            \label{fig:relat_diff_opal_oplib}
        \end{figure}

        Since we evolved these SCMs with the same input physics (see Sect.~\ref{section:solar_model}), the disagreements between models for a different mixture arise from differences in the choice of the opacity source and the effect of the solar calibration (different $Y_{0}$, $Z_{0}$, $\alpha_{\rm{MLT}}$; see Table~\ref{tab:optimal}).
        Figure~\ref{fig:relat_diff_opal_oplib} illustrates some differences on the RMO profile $\kappa_{R}$ between solar models computed with OPLIB$_{T}^{\rm{AAG21}}$, OP$_{T}^{\rm{AAG21}}$, and OPAL$_{T}^{\rm{AAG21}}$ for the AAG21 solar mixture at temperatures above $10^{4}$ K. At first glance, these models have similar RMOs (top panel), but their differences (middle panel) show that the OPLIB$_{T}^{\rm{AAG21}}$ opacity is higher than the OPAL$_{T}^{\rm{AAG21}}$ and OP$_{T}^{\rm{AAG21}}$ opacity for $\log T\in [4.8, 5.5]$, while for temperatures above $\log T > 5.5$ and $\log T > 6.3$, the OPLIB$_{T}^{\rm{AAG21}}$ opacity is almost systematically lower than the OPAL$_{T}^{\rm{AAG21}}$ and OP$_{T}^{\rm{AAG21}}$ opacity, respectively. The differences that we obtained between OPLIB$^{\rm{AAG21}}_{T}$ and OPAL$^{\rm{AAG21}}_{T}$ (green line in Fig.~\ref{fig:relat_diff_opal_oplib}) were reportted in previous studies, in particular, the large difference at low temperature $\log T\simeq4.1$ (see Fig.3 in \citealt{boudreaux2023updated}, Fig.5 in \citealt{farag2024expanded}).
        \paragraph{}
        By examining the relative differences between RMO computed from monochromatic data and those derived from pre-calculated tables (bottom panel of Fig.~\ref{fig:relat_diff_opal_oplib}), we see that unlike models using OP opacities, models based on OPLIB opacities exhibit larger discrepancies: around $\sim 5\%$ at temperatures above $\log T=4.5$, and up to approximately 15$\%$ at lower temperatures. In this specific low-temperature domain, the OPLIB$_{M}$ temperature grid is more sparsely populated than that of OPLIB$_{T}$ \citep{farag2024expanded}. The  difference peak is strongly linked to the interpolation on the temperature. Another way to explain these differences at low temperatures is the $\rho-\Ne$ correspondence. As previously mentioned, the Cesam2k20 user can choose how the electron density value is determined (see Sect.~\ref{section:interpolation}). We computed a new model (OPLIB$_{M}^{\rm{AAG21,}\rho}$) using the same internal structure as OPLIB$_{M}^{\rm{AAG21}}$, but using the ion fractions derived from the OPLIB database to compute $\Ne$ instead of from the Saha equation. The relative differences of the RMO profile between this model and OPLIB$_{T}^{\rm{AAG21}}$ model are shown in Fig.~\ref{fig:relat_diff_opal_oplib}. Compared with OPLIB$_{M}^{\rm{AAG21}}$, the differences at low temperatures are smaller than before, showing the effect of the choice of method for calculating ion fractions (and thus, $\Ne$) on the determination of the RMO in this context. This choice has no effect on the internal structure, however (see Appendix~\ref{appendix:calibration}). Finally, the effect of the calibration and the differences in the $T-\rho$ profile are negligible (see Appendix~\ref{appendix:calibration}).
    \subsection{Helioseismic results}
        
        In this section, we compare helioseismic quantities predicted by our SCMs with the helioseismic inference from \cite{basu2009}. These quantities are the helium mass fraction at the base of the convective zone and the corresponding radius ($r_{\rm{bcz}}$), and they are listed in Table~\ref{tab:properties}. 
        \begin{table}
        \caption{Properties of solar-calibrated models.}
        \label{tab:properties}
        \centering
        \renewcommand{\arraystretch}{1.35}
        \resizebox{0.5\textwidth}{!}{
        \begin{tabular}{l|ccc}
         \hline
         Properties & ($Z/X$)$_{\rm{surf}}$ & $Y_{\rm{bcz}}$ & $r_{\rm{bcz}}$~[$R_{\odot}$]\\
         \hline \hline
         OPAL$_T^{\text{AAG21}}$&0.0187 &0.2351 &0.724 \\
         OP$_T^{\text{AAG21}}$ &0.0187 &0.2343& 0.722 \\
         OP$_M^{\text{AAG21}}$ &0.0187 &0.2343 &0.723 \\
         OPLIB$_T^{\text{AAG21}}$ &0.0187&0.2234&0.721 \\
         OPLIB$_M^{\text{AAG21}}$ &0.0187&0.2287&0.721 \\
         \hline
         OPAL$_T^{\text{MB22}}$ &0.0225 &0.2445 &0.717 \\
         OP$_T^{\text{MB22}}$ &0.0225 &0.2440 &0.716\\
         OP$_M^{\text{MB22}}$ &0.0225 &0.2428 &0.715 \\
         OPLIB$_T^{\text{MB22}}$ &0.0225&0.2404&0.715 \\
         OPLIB$_M^{\text{MB22}}$ &0.0225&0.2383&0.713 \\
         \hline \hline
        Obs.\tablefootmark{a} &---& 0.2485 $\pm$ 0.0035& 0.713 $\pm$ 0.001 \\
         \hline
        \end{tabular}
        }
        \tablefoot{ \tablefoottext{a}{Helioseismic constraints on helium mass fraction $Y_{\rm{bcz}}$ and the radius of this zone $r_{\rm{bcz}}$ are taken from \cite{basu2009}}}
        \end{table}
        
        For AAG21 and MB22, the OPLIB$_{T}$ and especially OPLIB$_{M}$ solar-calibrated models produce a slightly deeper base of the convective zone than OPAL$_{T}$, OP$_{T}$, and OP$_{M}$ models, showing that the choice of the opacity source has a measurable effect on the $r_{\rm{bcz}}$ value. However, they also have a smaller helium mass fraction because their initial helium mass fraction is smaller (see Table~\ref{tab:properties}), showing that even though they improve the position of the base of the convective envelope, they do not resolve the solar modelling problem. By comparing OP$_{T}$ and OP$_{M}$ models for the two mixtures, smaller differences are seen between their helioseismic quantities relative to the OPLIB source. The greater differences between the OPLIB$_{T}^{\rm{AAG21}}$ and OPLIB$_{M}^{\rm{AAG21}}$ are visible for the AAg21 mixture, especially on the $Y_{\rm{bcz}}$ value. These differences are linked to the large difference in the $Y_{0}$ parameter between these models (see Table~\ref{tab:properties}).
        
        \begin{figure}
           \centering
           \includegraphics[width=\hsize]{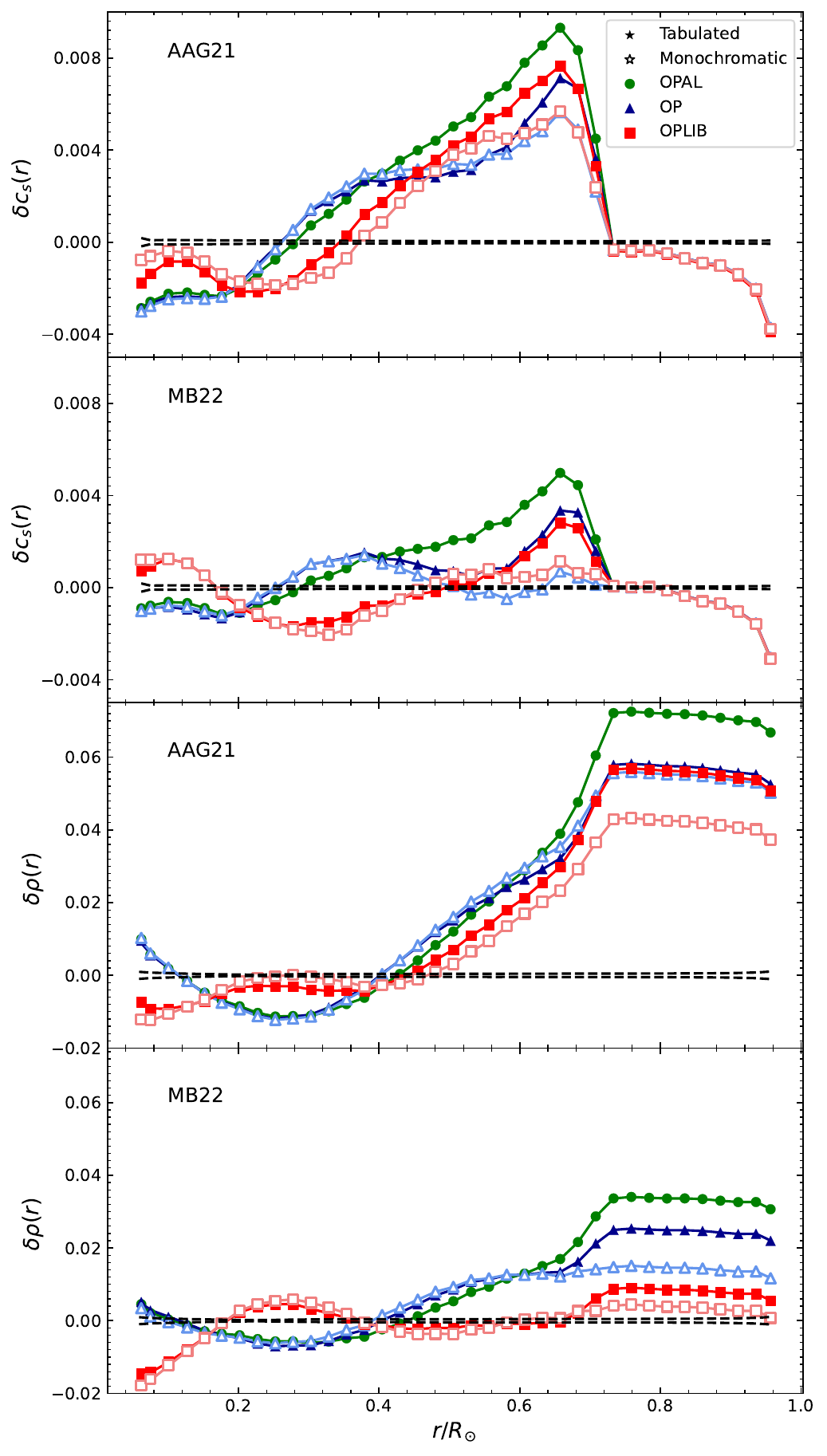}
           \caption{Fractional sound speed and density differences, $\delta c_{s}$ = ($c_{obs}$ - $c_{s} (r)$)/$c_{s} (r)$ and $\delta \rho$ = ($\rho_{obs}$ - $\rho(r)$)/$\rho(r)$, between the values predicted by an SCM, $c_{s} (r)$ and $\rho(r)$, and the $c_{obs}$ and $\rho_{obs}$ values inferred from helioseismic data \citep{basu2009} for the AAG21 and MB22 mixture. The 1$\sigma$ observational uncertainties are shown as the dark hashed lines at ordinates of zero. The coulored circles, triangles, and squares correspond to a different opacity source (OPAL, OP, and OPLIB, respectively). The full markers correspond to the model that includes RMO pre-calculated tables, and the outlined markers correspond to the model using monochromatic data. The markers show the locations where $\delta c_{s}$ and $\delta \rho$ are evaluated.}
           \label{fig:sound_speed}
        \end{figure}
        
        Figure~\ref{fig:sound_speed} shows the difference in the sound speed profiles $\delta c_{s}$ (top) and density profiles $\delta\rho$ (bottom) between the values predicted by these models and the values inferred from helioseismic data \citep{basu2009} for the AAG21 and MB22 mixtures for each SCM. The OPLIB$_{M}^{\rm{AAG21}}$ model agrees with OPLIB$_{T}^{\rm{AAG21}}$ in the stellar core for $r/R_{\odot} \leq 0.4$, but has significantly lower $\delta c_{s}$ and especially $\delta\rho$ than the other solar-calibrated models. For the AAG21 mixture, the calculated sound speed and density profiles using OPLIB pre-calculated opacity tables and especially monochromatic opacity data agree better with helioseismic constraints. For the MB22 mixture, the models using monochromatic opacity data OP$_{M}^{\rm{MB22}}$ and OPLIB$_{M}^{\rm{MB22}}$  have a consistent sound speed profile and density profile as the models using pre-calculated opacity tables (OP$_{T}^{\rm{MB22}}$ and OPLIB$_{T}^{\rm{MB22}}$ respectively) for $r/R_{\odot} \leq 0.6$, but have an improvement for $r/R_{\odot} > 0.6$. These improvements show that using monochromatic opacity data with the MB22 mixture strongly affects the density and sound speed profiles, especially around the base of the convective zone. For both mixtures, the SCMs using OPLIB opacities agree better overall in their density profile with helioseismic inference than the models with other opacities, although this indicator alone is not sufficient to assess the quality of a solar model. Overall, the MB22 models show the smallest $\delta c_{s}$ and $\delta \rho$ differences for all opacity sets, as was reported in previous studies \citep{magg2022,buldgen2024,farag2024expanded}.
    
    \subsection{Solar models with lithium depletion}
        \label{section:solar_model_Li}

        We also computed other solar-calibrated models using the lithium solar abundance as constraints (A(Li)\footnote{A(X) $\equiv \log(N_{X}/N_{H}) + 12$, with $N_{X}$ representing the number density of the element $X$.}$ = 0.96 \pm 0.05$ \citealt{wang2021lithium}). These new models used the same input physics as other models (see Sect.~\ref{sec:solar_model}) with an additional turbulent diffusion described by its coefficient $D_{\rm{turb}}$. This allowed us to reproduce the lithium abundance. We chose the prescription of \cite{richer2000} with $D_{\rm{turb}}$ defined as 
        \begin{equation}
        \label{eq:d_turb}
            D_{\rm{turb}}=\omega D(\rm{He})_{0}\left(\frac{\rho_{0}}{\rho}\right)^{n},
        \end{equation}
        where $\omega$ and $n$ are two constants defined by the Cesam2k20's user, and $D$(He) and $\rho$ are the local diffusion coefficient of helium and the mass density. The index 0 indicates that the value is taken at a reference envelope mass $\Delta M_{0}$ ($\Delta M \equiv M_{*} - M_{r}$ ; see \citealt{manchon2025} for more details). We set the constants $\omega$ and $n$ to $10^{4}$ and 4, respectively \citep{michaud2011horizontal}, and the value of the reference depth, $\Delta M_ {0}$, was adjusted to fit the lithium surface abundance of the Sun in addition to the other constraints listed in Table~\ref{tab:constraints}.
        \paragraph{}
        Table~\ref{tab:lithium_opti} regroups the calibrated parameters of these new SCMs. Compared to the values of the previous models (Table~\ref{tab:optimal}), we found that adding the turbulent diffusion reduces the mass fraction of the initial helium and metals for all new models, as expected. These differences affect the helium mass fraction at the base of the convective envelope. The helim mass fractions of OPLIB models are always lower than those in the OP models, and the convective envelope is also deeper. Similarly to Sect.~\ref{section:solar_model}, the model parameters based on pre-calculated opacity tables are more similar than those that were computed with monochromatic opacity data.
        
        \begin{table}
        \caption{Optimal parameters of SCMs including Li as a constraint.}
        \label{tab:lithium_opti}
        \centering
        \renewcommand{\arraystretch}{1.5}
        \resizebox{0.5\textwidth}{!}{
        \begin{tabular}{l|ccccc}
         \hline
         Models & $Y_{0}$ & $Z_{0}$ & ($Z/X$)$_{0}$ & $\alpha_{\rm{MLT}}$ & $\Delta M_{0}~[M_{\odot}]$\\
         \hline \hline
         OP$_{T\rm{,Li}}^{\rm{AAG21}}$&0.2572 &0.0147 & 0.0201 & 1.760918 & 0.018189 \\
         OP$_{M\rm{,Li}}^{\rm{AAG21}}$&0.2570 & 0.0147 & 0.0201 & 1.756517 & 0.018198\\
         OPLIB$_{T\rm{,Li}}^{\rm{AAG21}}$&0.2484 & 0.0148 & 0.0201 & 1.791815 & 0.019127\\
         OPLIB$_{M\rm{,Li}}^{\rm{AAG21}}$&0.2505 & 0.0148 & 0.0201 & 1.787768 & 0.019067\\
         \hline
         OP$_{T\rm{,Li}}^{\rm{MB22}}$&0.2677 & 0.0174 & 0.0243 & 1.787505 & 0.016696\\
         OP$_{M\rm{,Li}}^{\rm{MB22}}$&0.2665 & 0.0174 & 0.0243 & 1.791789 & 0.016659\\
         OPLIB$_{T\rm{,Li}}^{\rm{MB22}}$&0.2633 & 0.0174 & 0.0243 & 1.814561 & 0.017286\\
         OPLIB$_{M\rm{,Li}}^{\rm{MB22}}$&0.2608 & 0.0175 & 0.0242 & 1.820509 & 0.017606\\
         \hline
        \end{tabular}}
        \end{table}
        
        \paragraph{}
        Table~\ref{tab:lithium_properties} regroups the helium mass fraction at the base of the convective zone and the surface abundance of Li$^{7}$ and Be$^{9}$, predicted by our new SCMs. We compared them with the values from previous models (see Table.~\ref{tab:properties}) and with the bibliography \citep{basu2009,wang2021lithium,amarsi2024solar} and found that for both mixtures and both opacity sources, $Y_{\rm{bcz}}$ is higher and agrees better with the seismic value because the change in the initial helium mass fraction is related to the addition of turbulent diffusion. However, we note that these new models also have a thinner convective envelope, which disagrees more strongly with seismology (0.713 $\pm$ 0.001 \citealt{basu2009}) than previous models. Finally, we note that the models using the MB22 mixture obtain $Y_{\rm{bcz}}$ and $r_{\rm{bcz}}$ values that agree better with observations, but they fail to reproduce the Be surface abundance (1.21 $\pm$ 0.05 \citealt{amarsi2024solar}) because we did not adjust the power in Eq.~\ref{eq:d_turb}.
        \begin{table}
        \caption{Properties of SCMs using Li as constraint.}
        \label{tab:lithium_properties}
        \centering
        \renewcommand{\arraystretch}{1.5}
        \resizebox{0.5\textwidth}{!}{
        \begin{tabular}{l|ccccc}
         \hline
         Models & ($Z/X$)$_{\rm{surf}}$ & $Y_{\rm{bcz}}$ & $r_{\rm{bcz}}$ ($R_{\odot}$) & A(Li) & A(Be)\\
         \hline \hline
         OP$_{T\rm{,Li}}^{\rm{AAG21}}$&0.0187 & 0.2401 & 0.725 & 0.96 & 1.21\\
         OP$_{M\rm{,Li}}^{\rm{AAG21}}$&0.0187 & 0.2401 & 0.726 & 0.96 & 1.21\\
         OPLIB$_{T\rm{,Li}}^{\rm{AAG21}}$&0.0187 & 0.2323 & 0.724 & 0.96 & 1.21\\
         OPLIB$_{M\rm{,Li}}^{\rm{AAG21}}$&0.0187 & 0.2343 & 0.724 & 0.96 & 1.21\\
         \hline
         OP$_{T\rm{,Li}}^{\rm{MB22}}$&0.0225 & 0.2494 & 0.719 & 0.96 & 1.28\\
         OP$_{M\rm{,Li}}^{\rm{MB22}}$&0.0225 & 0.2484 & 0.719 & 0.96 & 1.28\\
         OPLIB$_{T\rm{,Li}}^{\rm{MB22}}$&0.0225 & 0.2456 & 0.717 & 0.96 & 1.28\\
         OPLIB$_{M\rm{,Li}}^{\rm{MB22}}$&0.0225 & 0.2434 & 0.717 & 0.96 & 1.28\\
         \hline \hline
         Observations &-& 0.2485\tablefootmark{a} & 0.713\tablefootmark{a} & 0.96\tablefootmark{b} & 1.21\tablefootmark{c} \\
         Uncertainty & -&  0.0035&  0.001 & 0.05 & 0.05\\
         \hline
        \end{tabular}
        }
        \tablefoot{\tablefoottext{a}{\cite{basu2009},\tablefoottext{b}{\cite{wang2021lithium}},\tablefoottext{c}{\cite{amarsi2024solar}} }}
        \end{table}
        \paragraph{}
        For both mixtures, the OPLIB$_{T,\rm{Li}}$ and OPLIB$_{M,\rm{Li}}$ solar-calibrated models have a slightly deeper base of the convective zone and a smaller helium mass fraction than OP$_{T,\rm{Li}}$ and OP$_{M,\rm{Li}}$ models, as before. However, the differences between their helioseismic values are much smaller than before. This reduction in the difference is also seen between the OP$_{T,\rm{Li}}$ and OP$_{M,\rm{Li}}$ models.
        
        \paragraph{}
        \begin{figure}
           \centering
           \includegraphics[width=0.95\hsize]{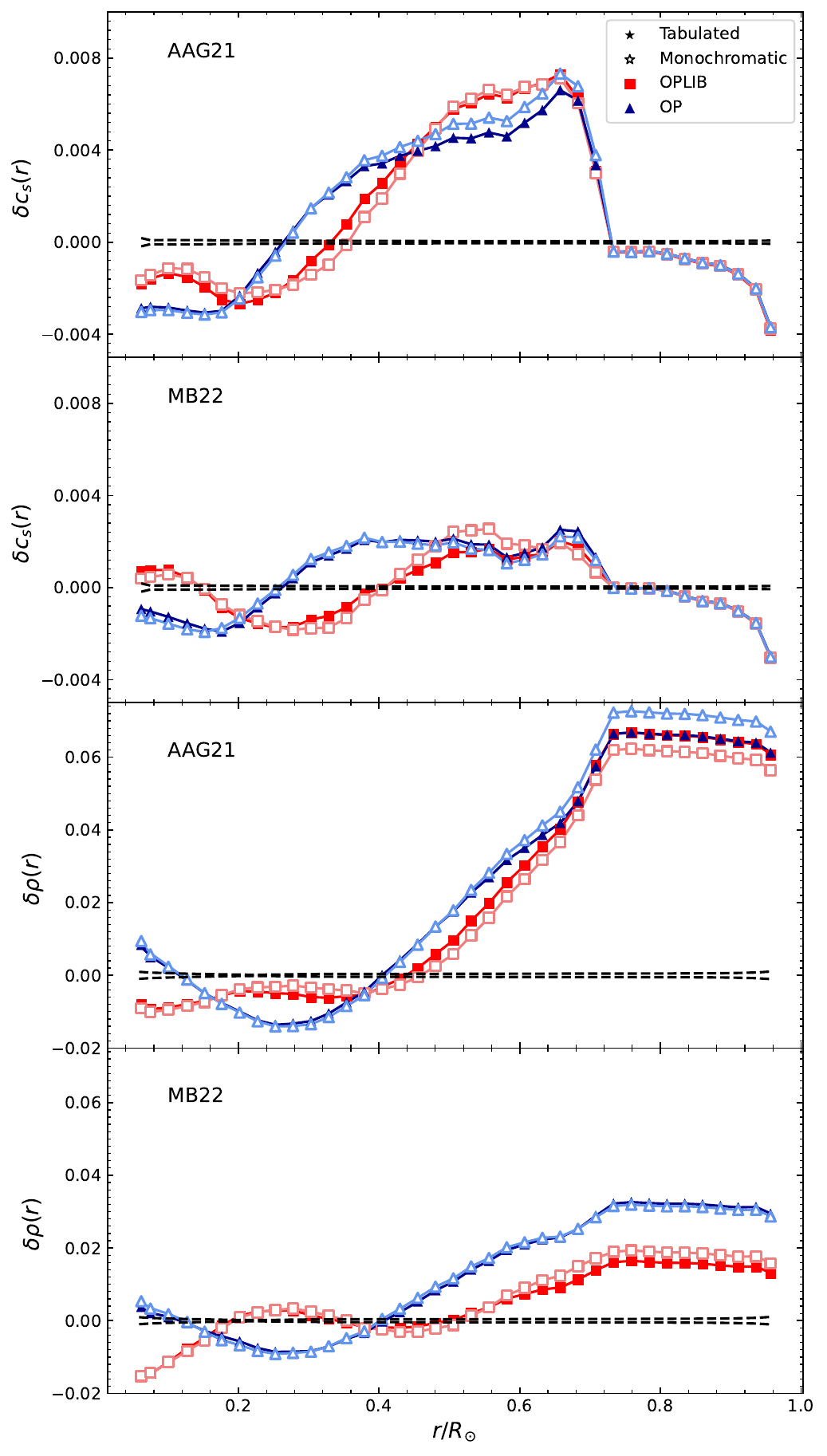}
           \caption{Same legend as Fig.~\ref{fig:sound_speed} for SCMs with lithium depletion.}
           \label{fig:sound_speed_Li}
        \end{figure}
        Figure~\ref{fig:sound_speed_Li} shows for each new SCMs including Li the difference in the sound speed profiles $\delta c_{s}$ (top) and density profiles $\delta\rho$ (bottom) between the values predicted by these models and the values inferred from helioseismic data \citep{basu2009} for the AAG21 and MB22 mixture. Some differences with previous profiles are clear, especially that the density profiles for all new models disagree more strongly with helioseismic constraints than before for $r/R_{\odot} \geq 0.5$. The OPLIB$_{T,\rm{Li}}$ and OPLIB$_{M,\rm{Li}}$ models always have the better density profiles for $r/R_{\odot} \geq 0.2$. For both mixtures, the sound speed profiles of all models, especially for the models computed with monochromatic opacity data (OP$_{M,\rm{Li}}$ and OPLIB$_{M,\rm{Li}}$), have more differences with helioseismic inferred ($c_{s}$ and $\rho$) than models without lithium calibration, similarly to the results found in the literature \citep{buldgen2025}. An important point to emphasise is that the sound speed profiles are now much more similar between models OP$_{T,\rm{Li}}$ and OP$_{M,\rm{Li}}$ and between models OPLIB$_{T,\rm{Li}}$ and OPLIB$_{M,\rm{Li}}$ for $r/R_{\odot} \geq 0.5$. Finally, the MB22 models show the smallest $\delta c_{s}$ and $\delta \rho$ differences for all opacity sets, which is again not sufficient to guarantee a good solar model \citep{buldgen2024}.
\section{Conclusion}
    \label{sec:conclusion}
    We retrieved the OPLIB monochromatic absorption opacty data for all elements, temperatures, and densities available in their website, and we implemented them in Cesam2k20 stellar evolution code. As for the OP monochromatic opacity data in Cesam2k20, we catalogued the OPLIB data in $T-\Ne$  grid for all pure elements, and it is therefore suitable for using monochromatic opacities \citep{huebner2014opacity}. We inferred the mass densities $\rho$ corresponding to our electron densities $\Ne$ grid using Eq.~\ref{eq:Ne_rho} and with a relative deviation $\Delta{\Ne/\Ne}\leq0.5\%$.
    \paragraph{}
    Following \cite{kilcrease2001plasma} and by using the method for OP monochromatic data \citep{boercker1987collective,seaton2005opacity}, we developed a new computation method for the total scattering opacity adapted for the OPLIB data. This method includes a more sophisticated collective effect for a mixture. The RMOs computed on our data grid with these new monochromatic opacity data and new scattering computation method agree very well with those of the OPLIB pre-calculated opacity tables from the database, which validates our implementation of these monochromatic opacity data in Cesam2k20.
    \paragraph{}
    Furthermore, by comparing the opacity profile of our calibrated models with OPLIB monochromatic opacity data and pre-calculated opacity tables available in the literature (OPAL \citealt{iglesias1996updated}, OP \citealt{seaton2005opacity}), we found the same results as in previous studies \citep{boudreaux2023updated,farag2024expanded}, which are that models using OPLIB pre-calculated opacity tables consistently predict a lower opacity in the solar core and higher opacity at the base of the convective envelope. We also found that there are more and larger differences in the RMO between OPLIB$_{T}^{\rm{AAG21}}$ and OPLIB$_{M}^{\rm{AAG21}}$ than between OP$_{T}^{\rm{AAG21}}$ and OP$_{M}^{\rm{AAG21}}$. In general, these differences are due to the interpolation on the temperature, especially at low temperature ($\log T \leq 4.5$). At these temperatures, the differences are also due to the method used to compute the electron density.
    \paragraph{}
    The solar-calibrated models generated by Cesam2k20 and using these new OPLIB monochromatic opacity data have a slightly deeper base of the convective zone and a smaller helium mass fraction than OPAL and OP for both mixtures. These comparisons are consistent with previous studies \citep{colgan2016new,boudreaux2023updated,farag2024expanded}. For both mixtures, the models generated with OPLIB opacities (pre-calculated and monochromatic) have a significantly lower initial mass fraction of helium than the models with OPAL and OP opacity, which strongly affects the value.
    \paragraph{}
    The use of monochromatic opacity data (OP or OPLIB) in solar modelling seems to improve the agreement with helioseismic constraints (sound speed and mass density profiles), especially with OPLIB monochromatic opacity data (see Fig.~\ref{fig:sound_speed}). However, with the models using the surface abundance of lithium as a constraint, the sound speed and mass density profiles of the models that include monochromatic opacity data are closer to those including pre-calculated opacity tables. This is due to the addition of turbulent diffusion, which reduces the relative variation in metals. Similarly to \cite{farag2024expanded}, we found that standard solar models computed with higher-metallicity mixtures (MB22) agree better with helioseismic constraints than low-metallicity mixtures (AAG21), regardless of the adopted opacity. However, a better agreement of the sound speed profile is not a sufficient criterion for the choice of a solar mixture \citep[see e.g.][]{buldgen2024}.
    \paragraph{}
    Finally, the use of monochromatic opacity data (including OP and OPLIB) offers numerous perspectives, particularly in the context of current and future helioseismology and asteroseismology missions such as CoRoT, Kepler, TESS, and in the future, PLATO. These missions need accurate stellar evolution models for the accurate characterisation of stellar systems in terms of mass, radius, and age. Stellar opacity plays an important role, especially in the determination of accurate stellar age, which can only be obtained from models. The accurate modelling of opacities in stellar evolution models is then essential to achieving the 10$\%$ PLATO/ESA requirement on the ages of stellar systems similar to the Solar System.
\begin{acknowledgements}
      We thank E. Boudreaux for her help with pyTOPScrap. We gratefully thank our anonymous referee whose comprehensive readings and remarks helped to improve the content of the manuscript. This work was supported by CNES, focused on PLATO. MV thanks to his partner for their support.
\end{acknowledgements}

\bibliographystyle{aa.bst}
\bibliography{aa61802-26}

\begin{appendix}
    
\section{Scattering computation method}
    \label{appendix:scatt}
    The OPLIB website also provides the scattering opacity, but since it depends on the mixture, it is computed at the moment request is made. This scattering opacity is the combination of Compton scattering and Rayleigh scattering \citep{colgan2016new}. To compute the Compton scattering, we followed the method adopted for OPLIB data presented by  \cite{kilcrease2001plasma} which defined the Compton scattering $\kappa_{\nu}^{\rm{compt}}$ as :
    \begin{equation}
        \rho\kappa_{\nu}^{\rm{compt}} = n_{e}\sigma_{T}G(u,T')R(\eta)f(\eta,\delta) 
        \label{eq:compton_scatt}
    \end{equation}
    with $n_{e}$ the free electron number density, $\sigma_{T}$ the natural Thomson scattering cross-section, $G(u,T')$ the Sampson’s relativistic correction to $\sigma_{T}$, $R(\eta,T')$ the Pauli blocking non-relativist factor and $f(\eta,\delta)$ collective effect. The dimensionless temperature $T'$ and $\delta$ parameter are defined as:
    \begin{equation}
        T' \equiv \frac{k_{B}T}{m_{e}c^{2}} \quad ; \quad \delta \equiv \frac{1}{2}\left(\frac{c}{2\pi\nu\lambda_{D}}\right)^{2}
    \end{equation}
    with $m_{e}$ the electron mass, $c$ the speed of light and $\lambda_{D}$ the Debye screening length.
    \paragraph{}
     We calculate the number density of free electron $n_{e}$ in Eq.\ref{eq:compton_scatt} by considering that bound electrons are free when the photon energy is equal or greater than the electron's binding energy \citep{colgan2016new}. With the information in the header of each OPLIB data, we can calculate this density only for hydrogen and helium.
    \paragraph{}
    The Sampson’s relativistic correction $G(u,T')$ is approximated at low temperatures and photon energy by :
    \begin{equation}
        \begin{split}
            G(u,T') = & 1 + 2T' + 5T'^{2} + \frac{15}{4}T'^{3} -\frac{1}{5}(16 + 103T' + 408 T'^{2})uT' \\ &
            + \left(\frac{21}{2} + \frac{609}{5}T' \right)(uT')^{2} - \frac{2203}{70}(uT')^{3}
        \end{split}
    \end{equation}
    In Cesam2k20, we decide to use this Equation only for temperatures $T$ lower than 2 keV while for higher temperatures, we directly interpolate in Sampson's tables \citep{sampson1959opacity} like for OPLIB tables \citep{colgan2016new}. The interpolation method used is bi-cubic spline interpolation.
    \paragraph{}
    The Pauli blocking non-relativistic factor $R(\eta)$ is a simple way to adjust the free electron density to account for the blocked states when electron degeneracy occurs \citep{kilcrease2001plasma} :
    \begin{equation}
        \label{eq:def_pauli}
        R(\eta)= \frac{\partial I_{1/2}(\eta)/\partial\eta}{I_{1/2}(\eta)} = \frac{I_{-1/2}(\eta)}{I_{1/2}(\eta)}
    \end{equation}
    where $I_{1/2}$ is the non-relativistic Fermi integral :
    \begin{equation}
        I_{s}(\eta) \equiv \frac{1}{\Gamma(s+1)}\int_{0}^{\infty}\frac{x^{s}}{e^{x-\eta}+1}dx \quad \text{for s $>$-1}
        \label{eq:int_fd}
    \end{equation}
    Unlike the scattering opacities of OPLIB, we excluded the relativistic correction of this factor since we are using some routines implemented in Cesam2k20 adapted for OP monochromatic data \citep{boercker1987collective} to reduce the numerical difference between the OP and OPLIB method. Another necessary clarification is that, with the OP procedure we determine the electron degeneracy parameter $\eta$ by using the relationship with the total electron density in the non-relativistic case:
    \begin{equation}
        \label{eq:eta}
        \Ne = \frac{2}{\lambda_{\rm{T}}^{3}}I_{1/2}(\eta)
    \end{equation}
    where $\lambda_{\rm{T}}$ is the thermal wavelength ($\lambda_{\rm{T}}=\sqrt{2\pi\hbar/mk_{B}T}$).
    \paragraph{}    
    Finally, we computed the collective effective contribution $f(\eta,\delta)$ following \cite{boercker1987collective} :
    \begin{equation}
        \label{eq:collective_effect}
        f(\eta,\delta) = 1-\frac{3}{8}\delta\left( \frac{k_{e}}{k_{D}}\right)^{2} \left[ (\delta^{3} + 2\delta^{2} + 2\delta)\ln{\left(\frac{\delta}{2+\delta}\right)} + 2\delta^{2} + 2\delta + \frac{8}{3}\right]
    \end{equation}
    where $k_{D}^{2}$ and $k_{e}^{2}$ are defined as :
    \begin{equation}
        \label{eq:kdke}
        k_{D}^{2} = k_{e}^{2} + k_{i}^{2} \quad ; \quad k_{e}^{2} = R(\eta)\frac{4\pi e^{2}\Ne}{k_{B}T} \quad ; \quad k_{i}^{2} = \frac{4\pi e^{2}\sum_{i}Z_{i}^{2}n_{i}}{k_{B}T}
    \end{equation}
    for a plasma with electron number density $\Ne$, ions number density $n_{i}$ (cm$^{-3}$) and ions of charge $Z_{i}$.
    
    The factor $k_{e}^{2}/k_{D}^{2}$ in Eq.\ref{eq:collective_effect} can be also written as :
    \begin{equation}
        \left( \frac{k_{D}}{k_{e}}\right)^{2} =1+\left( \frac{k_{i}}{k_{e}}\right)^{2} =1+\frac{\sum_{i}Z_{i}^{2}n_{i}}{R(\eta)\sum_{i}Z_{i}n_{i}}=1+\frac{\overline{N}_{f}^{2}}{R(\eta)\overline{N}_{f}}
    \end{equation}
    where $\overline{N}_{f}$ is the average number of free electrons per ion for the mixture and $\overline{N}_{f}^{2}$ is the average of the square of the number of free electrons over the ion states of the mixture. Like \cite{kilcrease2001plasma}, we neglected the quantum mechanical exchange in our computation of collective effects $f(\eta,\delta)$.
    
    \paragraph{}
    We also add the Rayleigh scattering opacities for the hydrogen and the helium element, following the method presented in \cite{colgan2016new}. Without the population of the ground state $N$ (in cm$^{-3}$) of the neutral atom for the metals, we do not include this scattering for these elements.
\section{Other sources of differences in RMO profile comparisons}
    \label{appendix:calibration}
    The bottom panel of Fig.~\ref{fig:relat_diff_opal_oplib} compares the RMO profiles derived from the various SCMs to assess the impact of using pre-calculated tables versus monochromatic opacity data. However, there are other sources of differences that we detail below.
    \subsection{Effect of the calculation method of $\Ne$ on the RMO profile}
        As previously mentioned, for OPLIB monochromatic data, the Cesam2k20 code allows the electron density $\Ne$ to be determined using the ion fractions derived from the OPLIB database or from the Saha equation. At the bottom of Fig.~\ref{fig:relat_diff_opal_oplib}, it can be seen that using the $\Ne$ value derived from OPLIB (black curve) reduces the discrepancy at low temperatures with the tabulated OPLIB opacity, compared to using the $\Ne$ value derived from the Saha equation (red curve). This shows the impact of the choice of method for calculating ion fractions (and thus $\Ne$) on the determination of the RMO.
        \paragraph{}
        To take this further and measure the impact of this choice on the internal structure, we computed a new SCM, denoted OPLIB$_{M}^{\rm{AAG21, C-}\rho}$, using the same input physics described in Sect.~\ref{section:solar_model}, and using the ion fractions derived from the OPLIB database to compute $\Ne$. This model is computed for the AAG21 mixture and the OPLIB monochromatic data. We show in Fig.~\ref{fig:relat_diff_appendix} that at the age of the Sun, the RMO profile of this new model is very similar to that of the uncalibrated model (OPLIB$_{M}^{\rm{AAG21,}\rho}$, see Fig.~\ref{fig:relat_diff_opal_oplib}).
        \paragraph{}
        However, the optimal parameters found for this new solar-calibrated model are very close to that of the OPLIB$_{M}^{\rm{AAG21}}$ model (see Table~\ref{tab:optimal}) and the relative differences between these parameters do not exceed 0.02 $\%$ : $\Delta Y_{0}/Y_ {0} \sim 0.015 \%$ ; $\Delta Z_{0}/Z_{0} \sim 0.002 \%$ ; $\Delta \alpha_{0}/\alpha_{0} \sim -0.019 \%$. These differences show that the method used to calculate electron density has very little impact on solar calibration and, consequently, little impact on the internal structure. To illustrate this point, we calculated the relative differences between the models' sound speed profiles $\Delta c_{s}$, defined as follows:
        \begin{equation}
            \Delta c_{s} = \frac{c_{s}(\rm{OPLIB}_{M}^{\rm{AAG21}}) - c_{s}(\rm{OPLIB}_{M}^{\rm{AAG21, C-}\rho})} {c_{s}(\rm{OPLIB}_{M}^{\rm{AAG21}})}
        \end{equation}
        
        The average of these differences is $|\Delta c_{s}| \sim 1.55\times10^{-5}$, and the maximum value is approximately $|\Delta c_{s}| \sim 6.40\times10^{-5}$, clearly demonstrating that models OPLIB$_{M}^{\rm{AAG21}}$ and OPLIB$_{M}^{\rm{AAG21, C-}\rho}$ have very similar internal structures, except for the RMO profile. The choice of how $\Ne$ is computed has then no impact on solar models in this context.
    \subsection{Effect of the calibration on the RMO profile comparisons}
\begin{figure}
            \centering
            \includegraphics[width=1.\hsize]{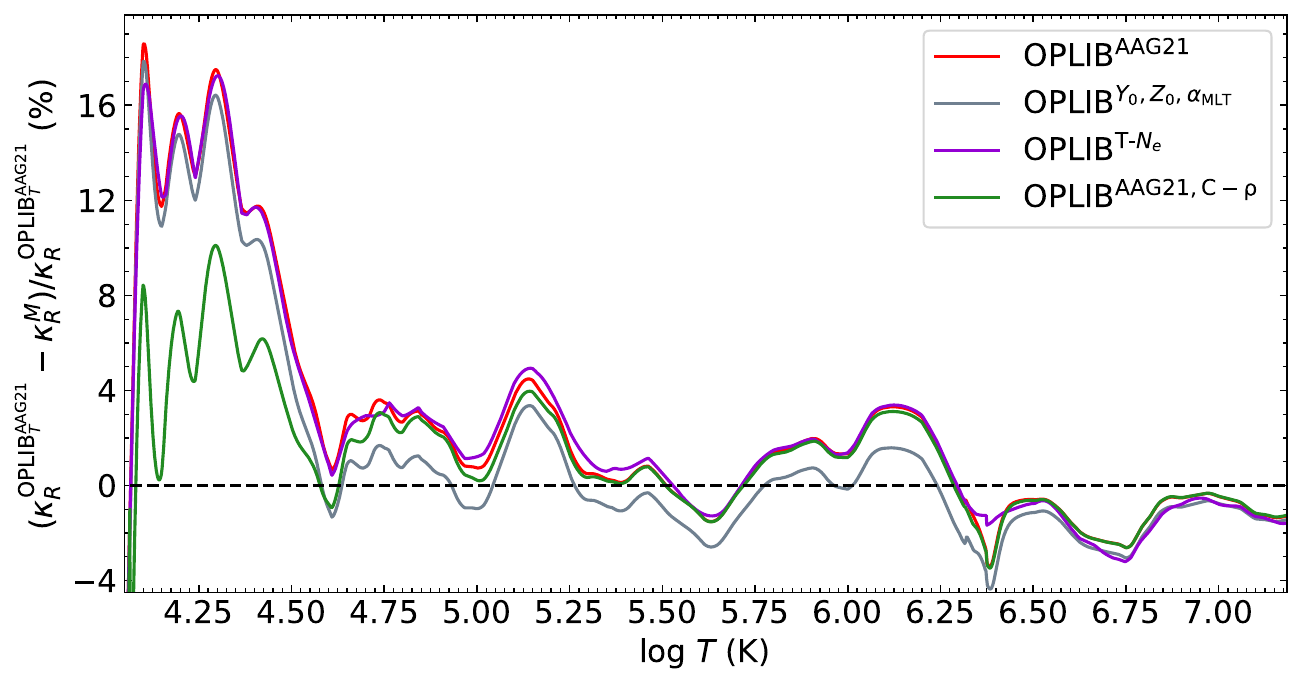}
            \caption{Relative differences of RMO between OPLIB$_{T}^{AAG21}$ model and  models using OPLIB monochromatic opacity data ($\kappa_{R}^{M}$). The red line corresponds to the SCM OPLIB$_{M}^{\rm{AAG21}}$ (same as in the Fig.~\ref{fig:relat_diff_opal_oplib}). The gray line represents the differences with the model using the initial values of $Y_{0}$, $Z_{0}$ and $\alpha_{\rm{MLT}}$ of the OPLIB$_{T}^{\rm{AAG21}}$ model. The purple line correspond to the model computed along the same $T-\Ne$ profile that the OPLIB$_{T}^{\rm{AAG21}}$ model. Finally, the green line (OPLIB$_{M}^{\rm{AAG21, C-}\rho}$) shows the difference with the new solar-calibrated model using the ion fractions derived from the OPLIB database to compute $\Ne$.}
            \label{fig:relat_diff_appendix}
        \end{figure}

        Table~\ref{tab:optimal} shows that the independently calibrated models have different $Y_{0}$, $Z_{0}$ and $\alpha_{\rm{MLT}}$. Consequently, the differences in Fig.~\ref{fig:relat_diff_opal_oplib} could reflect not only the opacity treatment itself, but also the response of the solar calibration to the changed opacity. To see these calibration effects, we computed a new model (OPLIB$_{M}^{Y_{0},Z_{0},\alpha_{\rm{MLT}}}$) using OPLIB monochromatic data and with the initial values of $Y_{0}$, $Z_{0}$ and $\alpha_{\rm{MLT}}$ of the OPLIB$_{T}^{\rm{AAG21}}$ model. At the present solar age, this new model exhibits a luminosity $L$, a radius $R$ and a photospheric abundance $(Z/X)_{\rm{surf}}$ similar to those of the OPLIB$_{M}^{\rm{AAG21}}$ model as confirmed by the relative differences between these values : $\Delta L/L_{\odot} = 2.205 \%$, $\Delta R/R_{\odot} = 0.474 \%$ and $\Delta (Z/X)_{\rm{surf}}/(Z/X)_{\rm{surf}}^{\rm{AAG21}} = 0.35 \%$. These results allow us to compare the opacity profile of this model with those of previous ones, since they share a relatively similar internal structure. The gray curve in Fig.~\ref{fig:relat_diff_appendix} shows the relative differences between the OPLIB$_{T}^{\rm{AAG21}}$ and OPLIB$_{M}^{Y_{0},Z_{0},\alpha_{\rm{MLT}}}$ models. Comparing with OPLIB$_{M}^{\rm{AAG21}}$, we see that calibration effects have a negligible impact on the comparison of profiles at high temperatures.
        \paragraph{}
        Finally, the models are not evaluated along the same temperature and electron density profile. The purple curve (OPLIB$_{M}^{\rm{T-}\Ne}$) in Fig.~\ref{fig:relat_diff_appendix} is the relative differences of the RMO between OPLIB$_{T}^{\rm{AAG21}}$ and a model using OPLIB monochromatic data along the same $T-\Ne$ profile as the OPLIB$_{T}^{\rm{AAG21}}$ model. These differences are very similar to OPLIB$_{M}^{\rm{AAG21}}$, where the maximal difference between these comparisons is around $\sim 2\%$ at $\log T = 3.37$.  We can conclude from these comparisons that  neither the calibration effect nor the difference in $T-\Ne$ profile has an impact on the comparison made in Section~\ref{sec:comparison_opacity}.

\end{appendix}
\end{document}